# Indigenous amino acids in primitive CR meteorites

Z. Martins[1]*, C. M. O'D. Alexander[2], G. E.Orzechowska[3], M. L. Fogel[4], P. Ehrenfreund[1]

[1]Astrobiology Laboratory, Leiden Institute of Chemistry, 2300 RA Leiden, The Netherlands
[2]Department of Terrestrial Magnetism, Carnegie Institution of Washington, Washington, D. C. 20015, USA.
[3]Jet Propulsion Laboratory, California Institute of Technology, Pasadena, CA 91109, USA.
[4]Geophysical Laboratory, Carnegie Institution of Washington, Washington, D. C. 20015, USA.

*Current address: Department of Earth Science and Engineering, Imperial College, London, SW7 2AZ, UK
E-mail: z.martins@imperial.ac.uk

## Abstract

CR chondrites are among the most primitive meteorites. In this paper, we report the first measurements of amino acids in Antarctic CR meteorites. Three CRs, EET92042, GRA95229 and GRO95577, were analyzed for their amino acid content using high performance liquid chromatography with UV fluorescence detection (HPLC-FD) and gas chromatography-mass spectrometry (GC-MS). Our data show that EET92042 and GRA95229 are the most amino acid-rich chondrites ever analyzed, with total amino acid concentrations ranging from 180 parts-per-million (ppm) to 249 ppm. The most abundant amino acids present in the EET92042 and GRA95229 meteorites are the α-amino acids glycine, isovaline, α-aminoisobutyric acid (α-AIB), and alanine, with $\delta^{13}C$ values ranging from +31.6‰ to +50.5‰. The carbon isotope results together with racemic enantiomeric ratios determined for most amino acids strongly indicate an extraterrestrial origin for these compounds. Compared to EET92042 and GRA95229, the more aqueously altered GRO95577 is depleted in amino acids. In both CRs and CMs, the absolute amino acid abundances appear to be related to the degree of aqueous alteration in their parent bodies. In addition, the relative abundances of α-AIB and β-alanine in the Antarctic CRs also appear to depend on the degree of aqueous alteration.

# Introduction

Meteorites provide crucial insights into the chemical processes occurring in the early solar system. In particular, the carbonaceous chondrite meteorites have a carbon-rich matrix, with some of its classes, the CM and CI chondrites, containing up to 2 wt% of organic carbon (for review see e.g. Sephton 2002). Meteorites have been investigated concerning their inventory of prebiotic molecules. Such compounds have properties (for example, chirality) that can be used to distinguish between terrestrial or extraterrestrial origins. Amino acids are therefore obvious candidates, and have been reported in several Antarctic and non-Antarctic meteorite samples (e.g. Cronin et al. 1979; Holzer and Oro 1979; Kotra et al. 1979; Shimoyama et al. 1979; Shimoyama and Harada 1984; Shimoyama et al. 1985; Botta and Bada 2002; Botta et al. 2002; Shimoyama and Ogasawara 2002; Glavin et al. 2006).

In previous work, the Antarctic Martian meteorites Elephant Moraine (EET) 79001 (McDonald and Bada 1995), Allan Hills (ALH) 84001 (Bada et al. 1998) and Miller Range (MIL) 03346 (Glavin et al. 2005) were analyzed for their amino acid content. In all three samples, the meteoritic amino acid distribution was similar to the one in the Allan Hills ice, which suggested that the ice meltwater was the source of the amino acids in these meteorites.

Antarctic micrometeorites (AMMs) have also been analyzed for the presence of amino acids (Briton et al. 1998; Glavin et al. 2004; Matrajt et al. 2004). The amino acids detected in most AMMs were present in low abundances, and showed a high L-enantiomeric excess, bearing similarities with those found in the Antarctic ice. To date, only one micrometeorite sample was found to contain α-AIB at significant levels (Brinton et al. 1998). Although the identification of α-AIB was tentative and needed further confirmation (Brinton et al. 1998), the concentration of α-AIB measured (~280 ppm) was higher than in any known meteorite.

Amino acids have also been reported in Antarctic carbonaceous chondrites showing different amino acid abundances. The CM2 ALH77306 (Cronin et al. 1979; Holzer and Oro 1979; Kotra et al. 1979), Yamato (Y-) 74662 (Shimoyama et al. 1979) and Lewis Cliff (LEW) 90500 (Botta and Bada 2002; Glavin et al. 2006) show an amino acid distribution and abundance similar to other non-Antarctic CM2 chondrites. However, other Antarctic CM2 chondrites, ALH83100 (Glavin et al. 2006), Y79331 and Belgica (B-) 7904 (Shimoyama and Harada 1984), contain lower quantities of amino acids. Last, the CM2 meteorite Y791198 has the highest concentration of amino acids (71 ppm) previously reported for a carbonaceous chondrite (Shimoyama et al. 1985; Shimoyama and Ogasawara 2002).

Several non-Antarctic carbonaceous chondrite meteorites have also been analyzed for amino acids (see e.g. Botta et al. 2002 and references therein), namely, the CM meteorites Murchison, Murray, Nogoya, Mighei and Essebi, which contain highly variable total amino acid abundances. Amino acid concentrations range from about 15 ppm for Murchison to about 6 ppm for Mighei. The CI1 chondrites Orgueil and Ivuna contain a much lower amino acid content, with total amino acid abundances of about 4.2 ppm (Ehrenfreund et al. 2001). The CV3 Allende and the ungrouped C2 Tagish Lake meteorites are found to be essentially free of amino acids. The trace amounts of amino acids that were detected are thought to be terrestrial contamination (Pizzarello e al. 2001; Botta et al. 2002; Kminek et al. 2002).

The CR chondrites are thought to contain the most primitive meteoritic insoluble organic material (see e.g. Cody and Alexander 2005). The CR2 Renazzo meteorite is the only reported fall in the CR group. To our knowledge this meteorite is also the only CR chondrite analyzed for amino acids. Renazzo has a total amino acid abundance that is similar to the CI chondrites Orgueil and Ivuna (Botta et al. 2002).

In the present paper we analyzed the amino acid content of two aqueously altered Antarctic CR2 chondrites: EET92042 and Graves Nunataks (GRA) 95229 (Grossman and Score 1996; Grossman 1998). A third sample, Grosvenor Mountains (GRO) 95577, is more aqueously altered than any other CR chondrite, and has been classified as the first CR1 by Weisberg and Prinz (2000). We have measured the amino acid abundances of these three meteorites by high performance liquid chromatography with UV fluorescence detection (HPLC-FD) and gas chromatography-mass spectrometry (GC-MS). Additionally, $\delta^{13}C$ values for most of the individual amino acids from the EET92042 and GRA95229 meteorites were obtained by gas chromatography-combustion-isotope ratio mass spectrometry (GC-C-IRMS).

## Materials and Methods

### Tools and chemicals

All the tools, ceramics and glassware used for sample processing were cleaned for organic contaminants by heating in aluminium foil at 500°C for 3 h. All tips and Eppendorf tubes were supplied sterilized by Sigma-Aldrich. Unless stated otherwise, all chemicals were obtained in high purity from Sigma-Aldrich. Ammonium hydroxide (28-30 wt%) and isovaline were purchased from Acros Organics. Methanol (absolute HPLC) was obtained from Biosolve Ltd. Sodium hydroxide and hydrochloric acid (37%) were acquired from Boom. AG® 50W-X8 cation exchange resin (100-200 mesh) was purchased from Bio-Rad.

### Meteorite sample preparation and amino acid extraction procedure

The Antarctic CR EET92042 was collected in the 1992 Antarctic Search for Meteorites (ANSMET) expedition, and both Antarctic CRs GRA95229 and GRO95577 in the 1995 field season. Chips of EET92042, GRA95229 and GRO95577 were provided by the Antarctic meteorite curator at the NASA Johnson Space Center, Houston. Each meteorite sample was separately crushed and homogenized into powder in a glove box with a flow of ultra high purity argon, using a ceramic mortar and pestle, and stored in sterilized glass vials. A serpentine sample provided by the Natural History Museum in Bern was grounded into powder in the same glove box, heated to 500°C for 3 h prior to being subjected to the same processing procedure as the meteorite samples and was used as a control blank.

Two separate sets of approx. 100 mg of each powdered meteorite and serpentine control blank samples were analyzed using the established procedure for extracting and analyzing amino acids in meteorites (Glavin et al. 2006; Botta et al. 2002; Zhao and Bada 1995). Both sets (sets 1 and 2) contained the EET92042, GRA95229 and GRO95577 meteorites, plus a procedural blank. Each of the samples, together with 1 ml of water, were flame sealed inside a test tube and heated for 24 h in a heating block set at 100°C. One of two equal parts of the water supernatants was then dried under vacuum and subjected to 6N acid vapor hydrolysis for 3 h at 150°C. The non-hydrolyzed extracts of the meteorite samples were not analyzed in this study. The acid hydrolyzed extracts of the samples were each brought up in 3 ml of HPLC water and then desalted on a cation exchange resin. The amino acids were eluted from the resin with 5 ml of ammonium hydroxide and the eluates were dried under vacuum. The residues were dissolved in 100 μl of water prior to analysis. Aliquots of sample set 1 were derivatized with *o*-phthaldialdehyde/N-acetyl-L-cysteine (OPA/NAC) and analyzed by HPLC-FD (based on the methods by Glavin et al. 2006; Botta et al. 2002). The remaining aliquots of sample set 1 were derivatized with trifluoroacetic anhydride (TFAA)/isopropanol and analyzed by GC-MS (based on the method by Pizzarello et al. 2004). A 10 μl aliquot of sample set 2 was also derivatized with *o*-phthaldialdehyde/N-acetyl-L-cysteine (OPA/NAC) and analyzed by HPLC-FD. The remaining portion of sample set 2 was derivatized with (TFAA)/isopropanol and analyzed by GC-C-IRMS (based on the method by Pizzarello et al. 2004).

**HPLC-FD analysis**

10 μl of 0.1 M sodium borate buffer were added to 10 μl aliquots of each sample extract (sets 1 and 2) present in Eppendorf vials. These were dried under vacuum to remove any residual ammonia, brought up in 20 μl of sodium borate buffer, and then derivatized with 5 μl of OPA/NAC. The derivatization was quenched after 1 or 15 min by adding 475 μl of 50mM sodium acetate buffer.

Separation by HPLC-FD of the OPA/NAC-derivatized amino acids was achieved in a C18 reverse phase (250 x 4.6 mm) Synergi 4μ Hydro-RP 80A column (from Phenomenex®) kept at room temperature, elution at 1 ml/min, using 50mM sodium acetate (4% methanol (v/v)) as buffer A, and methanol as buffer B. The gradient was 0 to 4 min, 0% buffer B; 4 to 5 min.0 to 20% buffer B; 5 to 10 min, 20% buffer B; 10 to 17 min, 20 to 30% buffer B; 17 to 27 min, 30 to 50% buffer B; 27 to 37 min, 60% buffer B; 37 to 49 min, 60% buffer B; 49 to 50 min, 60 to 0% buffer B; 50 to 60 min, 0% buffer B. UV fluorescence detection was performed on a Shimadzu RF-10A$_{XL}$ (excitation wavelength at 340 nm and emission at 450 nm). Amino acids were identified by retention time comparison with known standards (see Fig. 1). Amino acid abundances (part per billion by weight) were calculated by comparison to the integrated peak area of each sample, corrected for the abundances in the serpentine blank sample, with the integrated peak area of known amino acid standards. The calculated amino acid concentrations (see Table 1) are the average of five independent analyses of sample sets 1 and 2 for both 1 min and 15 min derivatization.

**GC-MS analysis**

Aliquots of each sample extract (set 1) were separately placed in 1 ml conical vials. The vials were placed under a stream of dry N$_2$ (60-80 ml/min) to evaporate water. For esterification, 100 μl of acetylchloride: isopropanol mixture (30:70 v/v) was added and the vials tightly capped with a Teflon-lined screw caps. Samples were placed in standard heating blocks for 1 h at 110°C. After cooling to room temperature the reagents excess was evaporated under the stream of dry $N_2$. 100 μl of methylene chloride and 50 μl of TFAA were added. The vials were tightly capped and heated at 100°C for 10 min. After the vials had cooled to room temperature, the excess reagent was removed under a stream of dry $N_2$. Finally, the derivatized samples were dissolved in 55 μl of ethyl acetate containing 18.3 ng/μl of pyrene, which was used as the external standard. 1 μl of sample was injected into the GC/FID/MS. GC-MS analyses were performed using a Varian Model GC-3800/FID/Ion-Trap Mass Spectrometer-Saturn 2000 equipped with an Electronic Pressure Control (EPC) system, and an autosampler Model 8200 (Varian). Injections of sample were performed using the autosampler programmed with a solvent flush sampling and a solvent plug of 0.2 μl, upper and lower air gaps, an injection rate of 0.2 μl/sec and a vial needle depth of 90%.

Separation of the D, L-amino acid enantiomers was achieved using a Helifex Chirasil-Val column (50 m x 0.25 mm ID x 16 μm film thickness) from Alltech. The end of the column was mounted into a Valco TEE connector, which splits the sample via transfer lines of 0.4 m x 0.1 mm ID and 1.6 m x 0.32 mm ID to the MS and FID, respectively. A very good alignment of corresponding peaks between the FID and the MS chromatograms, with a constant 0.08 min offset, was obtained. Helium was used as carrier gas with a flow of 2.3 ml/min. The injection port was set at 220ºC. The oven program was held for 5 min at 70ºC, increased by 2ºC/min to 100ºC, then increased to 200ºC by 4ºC/min and held for 30 min, and finally increased by 10ºC/min to 225ºC and hold for 5 min. Amino acids present in the meteorite samples were identified by comparison of the retention time and mass fragmentation pattern with known amino acid standard mixtures (see Fig. 2).

**GC-C-IRMS analysis**

Each extract of sample set 2 was derivatized separately using (TFAA)/isopropanol, and generally carried through the same procedure as described for the GC-MS analysis. The only differences were in the volumes of reagent used, that is, in the esterification step 500 μl of acetylchloride: isopropanol mixture were added to the samples, and on the next step 500 μl of methylene chloride and 500 μl of TFAA were used. Additionally, the Chirasil-Val column had the dimensions of 50 m x 0.32 mm ID (0.2 μm film thickness), and helium was used at a constant pressure of 15 PSI. Carrier gas and temperature program were the same as the GC-MS analysis. Amino acids were separated by the GC column, and then oxidised to $CO_2$ through the oxidation oven maintained at 980ºC. A Thermo Finnigan MAT Delta Plus-XL GC-C-IRMS was used to perform the carbon isotope analyses. $CO_2$ reference gas ($\delta^{13}C$ value of -41.10‰ PDB) was injected via the interface to the IRMS for the computation of $\delta^{13}C$ values of samples. Mixtures of amino acid standards were subjected to the entire TFAA/isopropanol derivatization procedure described before. The mixtures were run daily on the GC-C-IRMS, with typical standard deviation of ±0.99‰.

Carbon isotopic values were obtained by mass balance by measuring a set of standards (O'Brien et al. 2002): $\delta^{13}C$ amino acid standard derivatized = (% of carbon amino acid) (EA amino acid standard) + (% of carbon TFAA/isopropanol) ($\delta^{13}C$ TFAA/isopropanol), where the EA amino acid standard value is the $\delta^{13}C$ value of the amino acid standard established by a Carlo Erba elemental analyser (EA)-IRMS. Finally, the $\delta^{13}C$ values of the amino acids present in the meteorite samples were obtained by correcting for carbon added from the TFAA/isopropanol, and were calculated by mass balance: $\delta^{13}C$ amino acid in sample derivatized = (% of carbon in amino acid) ($\delta^{13}C$ amino acid in sample) + (% of carbon in TFAA/isopropanol) ($\delta^{13}C$ TFAA/isopropanol).

## Results

Fig. 1 displays typical HPLC-FD chromatograms of the acid hydrolyzed, hot-water extracts of the Antarctic CR meteorites plus a serpentine blank. The amino acid concentrations, determined by HPLC-FD, for EET92042, GRA95229 and GRO95577 are the average of several independent analyses of two different extracts (sets 1 and 2; see section *HPLC-FD analysis* for more details). The most abundant amino acids in the EET92042 and GRA95229 meteorites are glycine, D-alanine, L-alanine, α-AIB and isovaline (Table 1). Lower levels of valine, glutamic acid, β-amino-*n*-butyric acid (β-ABA), β-alanine, γ-amino-*n*-butyric acid (γ-ABA), β-aminoisobutyric acid (β-AIB) and aspartic acid were also present in both meteorites (Table 1).

The GRO95577 meteorite had the lowest concentration of amino acids, with values ranging from 8 ppb to 136 ppb (Table 1).

We further analyzed the three Antarctic CRs for amino acids using GC-MS in order to detect amino acids by their characteristic mass fragmentation patterns. The amino acid contents of the GRO95577 meteorite were below the GC-MS detection limits (~1 pmol). Fig. 2 shows a typical ion chromatogram of the acid hydrolyzed, hot-water extracts of the EET92042 and GRA95229 meteorites. All the detected amino acids and corresponding abundances are given in Table 2. The GC-MS analysis confirmed the results obtained by HPLC-FD, with values generally agreeing within the associated errors, or at least in the same order of magnitude. The most abundant amino acids for both CR2 chondrites matched those determined by HPLC-FD.

The non-hydrolyzed (free) extracts of the three Antarctic CR meteorites were not analyzed in this paper. Analysis of the non-hydrolyzed extract of GRA95229 has recently been performed by Pizzarello and Garvie (2007). The results for the few amino acids analyzed show that hydrolysis of the meteorite extract yielded only a small increase on the amino acid abundance.

## Discussion

**Indigenous and terrestrial amino acids**

The EET92042 and GRA95229 meteorites have the highest amino acid contents ever measured in any carbonaceous chondrite (Tables 1 and 2). The total amino acid abundances in these CR2 chondrites, 180 ppm and 249 ppm, for EET92042 and GRA95229 respectively (Table 1), are at least a factor 10 higher than almost all other primitive chondrites, such as the CM2s Murchison and Murray (e.g. Ehrenfreund et al. 2001).

The total amino acid concentrations of EET92042 and GRA95229 are only comparable to the total amino acid abundance of the Antarctic CM2 Y791198, but are still at least a factor of 2.5 higher (Shimoyama and Ogasawara 2002). If the amino acids detected in EET92042 and GRA95229 (or their precursors) were formed prior to accretion of the meteorite parent body (or bodies), they would not survive chondrule and CAI formation. The amino acids (or their precursors) must, therefore, have been accreted with the meteorite matrices. If this was the case, then the appropriate comparison should be between amino acid matrix-normalized abundances. As CMs contain more than 50 vol% matrix (McSween 1979) and CRs about 30 vol% matrix (Weisberg et al. 1993), this would imply that the amino acid matrix-normalized concentrations would be even higher in EET92042 and GRA95229 than in CMs. Fig. 3A displays the total amino acid abundances (in ppb) for the CM2 Y791198 (Shimoyama et al. 1985; Shimoyama and Ogasawara 2002), the CM1 LAP02277 (Botta et al. 2007), the CR2s Renazzo (Botta et al. 2002), EET92042 and GRA95229 (this work, Table 1 and 2), and the CR1 GRO95577 (this work, Table 1), while Fig. 3B displays the relative amino acid abundances (glycine = 1) for the same meteorites. The amino acid distribution in Y791198 looks similar to the Antarctic CR2 meteorites EET92042 and GRA95229. This may indicate that these three meteorites had similar amino acid precursor material available.

Previous analysis of AMMs revealed a high abundance of α-AIB (~280 ppm) in one micrometeorite sample (Briton et al. 1998). Although this result needs further confirmation (Brinton et al. 1998) and most AMMs have very low amino acid content (Briton et al. 1998; Glavin et al. 2004; Matrajt et al. 2004), there is the possibility that some micrometeorites have amino acid concentrations similar to the Antarctic CR2s analyzed in this study.

As the chondrites with the highest amino acid concentrations are from Antarctica, we must consider the possibility of terrestrial contamination. We used four approaches to determine whether the amino acids, present in the Antarctic CRs EET92042, GRA95229 and GRO95577 are terrestrial or extraterrestrial. These were: (*i*) detection of amino acids that are unusual in the terrestrial environment, (*ii*) comparison of the absolute abundances of amino acids in the meteorites to the levels found in the fall environment, (*iii*) determination of enantiomeric ratios, and (*iv*) measurement of the compound specific carbon isotopic compositions.

*The presence of amino acids that are rare in terrestrial proteins*

The amino acids α-AIB, isovaline, β-ABA and β-AIB were detected in the EET92042 and GRA95229 meteorites using HPLC-FD (Table 1) and GC-MS (Table 2). The GRO95577 meteorite also contained α-AIB, isovaline, β-ABA and β-AIB, but in low abundances (Table 1). These amino acids have also been detected in the CM2 Antarctic meteorites ALH77306 (Cronin et al. 1979; Kotra et al. 1979), Y74662 (Shimoyama et al. 1979), LEW90500 (Botta and Bada 2002; Glavin et al. 2006), ALH83100 (Glavin et al. 2006) and Y791198 (Shimoyama et al. 1985; Shimoyama and Ogasawara 2002). Except for Y791198, all these amino acids were present in much lower abundances than in EET92042 and GRA95229, with concentrations usually on the order of ~100 ppb (Cronin et al. 1979; Kotra et al. 1979; Botta and Bada 2002; Glavin et al. 2006). LEW90500 (Glavin et al. 2006) contained a slightly higher abundance of α-AIB (2706 ppb) and isovaline (1306 ppb). Y791198 (Shimoyama and Ogasawara 2002) contained similar abundances of α-AIB (22630 ppb), β-ABA (< 3250 ppb) and β-AIB (1835 ppb) as EET92042 and GRA95229, but lower abundance of isovaline (4075 ppb).

*Amino acid content of the meteorite fall site*

The potential for contamination from the surrounding environment includes ice and microbial biomass, and it is important for us to consider these sources. To our knowledge, ice from the Elephant Moraine (EET), Graves Nunataks (GRA) or Grosvenor Mountains (GRO) Antarctic regions has not been analyzed for amino acids. However, amino acid analyses of Allan Hills (McDonald and Bada 1995; Bada et al. 1998) and La Paz Antarctic ices (Glavin et al. 2006) showed similar distributions, with trace levels of aspartic acid, serine, glycine and alanine (1 ppb of total amino acid concentration). No isovaline, β-ABA or β-AIB was detected above detection limits. Only an upper limit of α-AIB (<2 parts-per-trillion (ppt)) was detected in the Allan Hills ice (Bada et al. 1998), while a relatively high abundance (46 ppt) of α-AIB was detected in a La Paz Antarctic ice sample (Glavin et al. 2006). However, these concentrations are $10^6$ times lower than the α-AIB values found in the EET92042 and GRA95229 meteorites, and $10^3$ times lower than values measured for GRO95577. Most likely the Antarctic ice was not the source of α-AIB, isovaline, β-ABA and β-AIB detected in EET92042, GRA95229 and GRO95577.

*D/L enantiomeric ratios*

The amino acid enantiomeric ratios (Table 3) for both protein and non-protein amino acids in EET92042 and GRA95229 are nearly racemic (D/L ~ 1), indicating either an abiotic synthetic origin, very long terrestrial residence ages or elevated temperatures at some point in their histories. The only exception is glutamic acid, which showed L-enantiomeric excesses (Table 3). The D/L ratio for glutamic acid in EET92042 (0.58 measured by HPLC-FD and 0.69 by GC-MS) and in GRA95229 (0.82 measured by HPLC-FD and 0.83 by GC-MS) can be explained by terrestrial L-glutamic acid contamination of the meteorites during their residence time on Earth. Biologically derived glutamic acid is principally in the L-form, therefore any addition of terrestrial glutamic acid would decrease the D/L. Possible sources of glutamic acid contamination include the meteorite fall site, i.e. Antarctic ice, and contamination during sample curation. Although glutamic acid was detected in Antarctic ices (McDonald and Bada 1995; Glavin et al. 2006), it was only present at residual levels (<0.3 ppb). Alternatively, terrestrial contamination during the curation history of the Antarctic CR2s may have contributed to the glutamic acid L-enantiomeric excess.

The amino acid enantiomeric ratios for the GRO95577 meteorite are all smaller than 0.8 (Table 3), which is an indication of significant terrestrial contamination.

Although the D- and L-enantiomers of isovaline are separated using HPLC-FD (Fig. 1), there is a possibility that α-ABA co-elutes with isovaline under the chromatographic conditions used (Glavin et al. 2006). This would explain the high peak area observed for L-isovaline relative to D-isovaline (respectively peaks 15 and 14; Fig. 1). Further analyses are currently being performed to test if α-ABA indeed interferes with isovaline in the fluorescence trace. If this is the case, then LC-MS analyses will be required for accurate isovaline enantiomeric measurements. Under the GC-MS conditions used in this study isovaline enantiomers could not be separated (Fig. 2), and therefore we cannot rule out potential sources of terrestrial contamination. Although isovaline is unusual in the terrestrial biosphere (see above), it may occur in bacteria and fungal peptides in the D-configuration (e.g., Keller et al. 1990). Therefore, if any microbes were present in the EET92042 and GRA95229 meteorites, these would increase the original D/L isovaline ratio. As we will discuss in the next paragraph, the GRA95229 meteorite has a high $\delta^{13}C$ value (+50.5‰) for isovaline, well outside the terrestrial range, providing compelling evidence for an extraterrestrial origin of this amino acid. Separation of the D- and L-enantiomers of isovaline by GC-MS is currently being carried out, and will be the subject of a future paper.

*Compound-specific carbon isotopic measurements*

We have focused our carbon isotope measurements on the most abundant amino acids present in the EET92042 and GRA95229 meteorites, which were the α-amino acids including glycine, alanine, α-AIB, and isovaline. We also analyzed the common biological amino acids, glutamic and aspartic acids, because these could be terrestrial contaminants. Amino acid abundances in GRO95577 were too low for carbon isotopic analysis (detection limits ~ 1 pmol).

The $\delta^{13}C$ values of α-amino acids present in the EET92042 meteorite ranged from +31.8‰ for glycine to +49.9‰ for L-alanine, while in the GRA95229 meteorite values ranged from +31.6‰ for α-AIB to +50.5‰ for isovaline (Fig. 4 and Table 4). These $\delta^{13}C$ values are clearly outside the terrestrial range (from -70.47‰ to +11.25‰) (Scott et al. 2006) and agree with the $\delta^{13}C$ values of the same α-amino acids (glycine, alanine, α-AIB, and isovaline) measured by other authors in the CM2 chondrite Murchison (Pizzarello et al. 2004). The similarity in $\delta^{13}C$ values may indicate a common reservoir (interstellar and protosolar) for the amino acid precursors in the CR2 and CM2 meteorites.

EET92042 shows $\delta^{13}C$ values for the L- and D-enantiomers of alanine that are similar (+49.9‰ and +44.5±2.0‰, respectively), which is in agreement with the D/L alanine ratio of ~1 seen before (Table 3). In the case of GRA95229, L-alanine and D-alanine have also high and identical $\delta^{13}C$ values (+40.9 ± 6.2‰ and +41.7 ± 2.4‰, respectively) within the associated errors, indicating that unless terrestrial contamination was limited to very specific peptides with equal amounts of D- and L-alanine, terrestrial contamination was minimal.

The carbon isotopic analysis of the glutamic acid showed that both meteorites have substantially lower $\delta^{13}C$ values for the L-enantiomer, even falling into the negative range (-19.5 ± 1.7‰ and -17.6 ± 1.9‰, respectively for EET92042 and GRA95229), while the D-enantiomer is rich in $^{13}C$ (+46.1 ± 2.1‰ and +47.2‰, respectively for EET92042 and GRA95229). This is consistent with the L-enantiomeric excess being due to terrestrial contamination described previously. However, the $\delta^{13}C$ values for L-glutamic acid are remarkably low compared to the $\delta^{13}C$ values for D-glutamic acid. Mass balance calculations provide some constrains on how this contamination may have occurred. A typical $\delta^{13}C$ composition for amino acids in the Antarctic environment is roughly -25‰, although the full terrestrial range is from -60.93‰ to -0.30‰ (Scott et al. 2006). If the indigenous extraterrestrial D- and L-enantiomers had the same isotopic compositions and the L-enantiomers were contaminated by terrestrial material with a $\delta^{13}C\approx$ -25‰, the indigenous material must have D/L ratios of about 8.9 and 8.1 for EET92042 and GRA95229, respectively. On the other hand, if one assumes that the indigenous material is racemic and that the D- and L-enantiomers have the same $\delta^{13}C$ composition, the heaviest isotopic compositions for the contaminants allowed by the abundance errors are $\delta^{13}C\approx$ -95‰ and -149‰ for EET92042 and GRA95229, respectively. These $\delta^{13}C$ values are well outside the known terrestrial range for amino acids (Scott et al. 2006). Since neither a non-racemic composition nor terrestrial contaminants with unusual compositions seem likely, a possible alternative explanation would be that most of the extraterrestrial L-glutamic acid was preferentially destroyed by microorganisms and terrestrial glutamic acid was subsequently added. However, to our knowledge microorganisms do not preferentially destroy L-glutamic acid. Additionally, microbial contamination would affect other amino acid, such as glycine, aspartic acid and alanine (Howe et al. 1965). The heavy $\delta^{13}C$ values for these amino acids present in EET92042 and GRA95229 (Table 4), as well as the racemic enantiomeric ratios for aspartic acid and alanine (Table 3), suggests that most of the glycine, aspartic acid and alanine are indigenous. At present, while it seems clear that L-enantiomers of glutamic acid are contaminated, but how it occurred remains unclear.

EET92042 has a $\delta^{13}C$ value for aspartic acid that is slightly lower for the L-enantiomer, while GRA95229 has $\delta^{13}C$ values for the aspartic acid enantiomers that are equivalent within the associated errors (Table 4). The stable carbon isotopic analysis showed that, except for L-glutamic acid, all the amino acids analyzed in this study and present in the EET92042 and GRA95229 meteorites are highly enriched in $^{13}C$, suggesting an extraterrestrial origin for the carbon in these compounds.

**Formation of α-meteoritic amino acids**

The EET92042 and GRA95229 meteorites are the most amino acid-rich carbonaceous chondrites reported to date. Racemic enantiomeric ratios, as well as the highly enriched $\delta^{13}C$ values, indicate primitive indigenous organic matter. These findings are supported by Busemann et al. (2006), who reported D and $^{15}N$ hotspots in EET92042 insoluble macromolecular organic matter, showing that primitive organic matter was preserved in this meteorite. Both meteorites have amino acid distributions, total amino concentrations, D/L enantiomeric ratios, and carbon isotope values of most individual amino acids that are very similar.

The high α-amino acid content (Table 1 and Table 2) is suggestive of a two-step formation process for these amino acids (Cronin et al. 1995), in which the amino acid precursors (aldehydes, ketones, ammonia and HCN) were present (or formed) in the protosolar nebula, and later incorporated into the asteroidal parent body. During aqueous alteration on the parent body, Strecker-cyanohydrin synthesis would have taken place to form the α-amino acids (Peltzer et al. 1984; Cronin and Chang 1993; Lerner et al. 1993). Since the carbonyl precursors (aldehydes and ketones) are thought to be synthesized by the addition of a one-carbon donor to the growing alkane chain, a decrease of the α-amino acid abundances (e.g. glycine > alanine > α-amino-*n*-butyric acid (α-ABA)) with increasing chain length would be expected. This trend was observed for example in the CM2 Murchison (e.g. Cronin and Chang 1993).

In the case of the EET92042 and GRA95229 meteorites, an exceptionally high alanine concentration is found, which does not follow the expected trend. If the α-amino acids were formed by Strecker synthesis, the high alanine abundances suggest a high abundance of the precursor acetaldehyde on the parent body (or bodies) of these meteorites. To our knowledge, EET92042 and GRA95229 were never analyzed for acetaldehyde (or other aldehydes and ketones). Future work should focus on the detection of aldehydes and ketones in these two meteorites to test whether Strecker-cyanohydrin synthesis is the correct formation mechanism for the α-amino acids detected.

Also, synthesis of branched carbon chain analogues is thought to be favored over straight carbon chain analogues (e.g. Cronin and Chang 1993). This trend is observed in the EET92042 and GRA95229 meteorites. For example, the abundance of the branched α-AIB in both of these meteorites is higher than the straight chain analogue α-ABA (Table 1 and Table 2).

Besides the high α-amino acid content, EET92042 and GRA95229 contain lower levels of other amino acids (Table 1 and 2). The abundances of non-α-amino acid are also higher in EET92042 and GRA95229 than in any other carbonaceous meteorite (Ehrenfreund et al. 2001; Botta et al. 2002). The non-α-amino acid cannot be produced by the Strecker-cyanohydrin synthesis, but are formed instead by other synthetic pathways. For example, β-amino acids are thought to be synthesized by Michael addition of ammonia to α,β-unsaturated nitriles, followed by hydrolysis (Cronin and Chang 1993). Additional synthetic pathways have been proposed for β- and γ-amino acids (for a review see e.g. Cronin and Chang 1993).

**Causes of amino acid abundance variation between meteorites**

The total amino acid abundances in EET92042 and GRA95229 are one to two orders of magnitude higher than in Renazzo (Botta et al. 2002) and GRO95577 (Fig. 3). Similar differences in abundance are seen between Y791198 and other CMs.

A possible reason for the low amino acid concentrations in GRO95577 is leaching of amino acids during its residence time in Antarctica. However, this would not explain the low abundances in Renazzo. Renazzo is a non-Antarctic CR2 that was quickly recovered after its fall. Therefore it is highly unlikely that this meteorite lost its amino acids as a result of weathering. Renazzo has a total amino acid concentration of only 4.8 ppm (Botta et al. 2002), much lower than the Antarctic CR2s analyzed here. Additionally, Renazzo has a distinct amino acid distribution, with γ-ABA (1092 ppb), glycine (875 ppb) and L-glutamic acid (856 ppb) as the most abundant amino acids (Botta et al. 2002). Only upper limits for alanine and α-AIB concentrations were reported for this meteorite, while isovaline was tentatively identified. Similarly, the CM2s Murchison and Murray were recovered soon after falling but have much lower amino acid abundances than the CM2 Y791198 (Ehrenfreund et al. 2001; Shimoyama and Ogasawara 2002).

Another possibility is that EET92042 and GRA95229, GRO95577 and Renazzo originated on at least three separate parent bodies. Amino acid formation may have been less active on the GRO95577 and Renazzo parent bodies due to a lack of amino acid precursors. However, there are no differences in the bulk petrologic or compositional properties of these meteorites that suggest they came from separate parent bodies (Weisberg et al. 1993; Grossman and Score 1996; Grossman 1998; Clayton and Mayeda 1999; Weisberg and Prinz 2000). The same arguments would lead to the conclusion that the CMs come from more than one parent body, but again there is no bulk compositional or petrologic evidence for this. While it cannot be ruled out, at present multiple parent bodies seems an unlikely explanation for the variations in amino acid abundance between meteorites.

Degradation or removal of the amino acids as a result of a higher degree of aqueous alteration might be another explanation for the differences among CR meteorites. Among the potential degradation/removal processes, oxidation may have been an important mechanism early in the aqueous alteration history (Cody and Alexander 2005). During aqueous alteration, low temperature chemical oxidation would have increasingly removed the aliphatic moieties in the free and macromolecular matter (Sephton et al. 2004; Cody and Alexander 2005; Martins et al. 2006). This trend is clearly seen for the amino acid content in the CM group, as the total amino acid abundances decrease from the least aqueously altered (Chizmadia and Brearley 2003) CM2 Y791198 (Shimoyama and Ogasawara 2002) to the more aqueously altered CM1s LAP02277 (Fig. 3), ALH88045 and MET01070 (Botta et al. 2007).

Except for GRO95577, Renazzo was shown to be generally more aqueously altered than the Antarctic CR meteorites (Weisberg et al. 1993). As pointed out by Glavin et al. (2006), the relative abundance of β-alanine (relative to glycine) appears to be generally higher in meteorites that have experienced more extensive aqueous alteration, while the relative abundance of α-AIB in these meteorites is lower than in the less aqueous altered meteorites. In Renazzo (Botta et al. 2002) the relative abundance of β-alanine (0.25; Fig. 3B) is higher than in EET92042 and GRA95229 (respectively 0.11 and 0.05; Table 1; Fig. 3B). Also, the relative abundance of α-AIB (Botta et al. 2002) in Renazzo is lower (<0.08; Fig. 3B) than in EET92042 and GRA95229 (respectively 2.15 and 0.48; Table 1; Fig. 3B). These results support the aqueous alteration hypothesis. While at present we have not identified the mechanism(s) for the variations in amino acid abundances between meteorites, parent body processes seem the most likely explanation.

## Conclusion

We have analyzed the amino acid content of three Antarctic CR meteorites, EET92042, GRA95229 and GRO95577. The total amino acid abundances in the CR2 chondrites EET92042 and GRA95229 were found to be the highest ever detected in any meteorite. This could be the result of CR chondrites being the most primitive and least aqueously altered meteorites. Compared to these two meteorites, the CR1 GRO95577 is depleted in amino acids. The CR2 meteorites EET92042 and GRA95229 have similar amino acid distributions to the CM2 Y791198. This fact, together with similar carbon isotope values for the amino acids present in the Antarctic CR2s and the CM2 Murchison may indicate a common reservoir in the interstellar medium and/or protosolar nebula for the amino acid or their precursors in both CR2s and CM2s.

The racemic enantiomeric ratios and the high $\delta^{13}C$ values determined for nearly all the amino acids present in the EET92042 and GRA95229 meteorites indicate that the compounds have a primarily extraterrestrial origin. The rich amino acid content observed in the EET92042 and GRA95229 meteorites make these Antarctic CR chondrites the most scientifically valuable of the carbonaceous meteorites. Further investigation of their carbonaceous inventory and other CR chondrite samples may help to reveal the processes which occurred in the early solar system that formed abundant organic prebiotic material.

## Acknowledgements

This research was supported by Fundação para a Ciência e a Tecnologia (scholarship SFRH/BD/10518/2002) and NWO-VI 016023003. C.M.O'D.A. was funded by NASA's Origins of Solar Systems program and Astrobiology Institute and M. L. F was supported by the NASA Astrobiology Institute through Cooperative Agreement NNA04CC09A. The authors would like to thank the Meteorite Working Group and Cecilia Satterwhite for providing the meteorite samples. We are grateful to D. Glavin and an anonymous reviewer for their constructive comments.
# References

Bada J. L., Glavin D. P., McDonald G. D. and Becker L. 1998. A search for endogenous amino acids in Martian meteorite ALH84001. *Science* 279: 362-365.

Botta O. and Bada J. L. 2002. Amino Acids in the Antarctic CM meteorite LEW90500 (abstract #1391). 33$^{rd}$ Lunar and Planetary Science Conference. CD-ROM.

Botta O., Glavin D. P., Kminek G., and Bada J. L. 2002. Relative amino acid concentrations as a signature for parent body processes of carbonaceous chondrites. *Origins of Life and Evolution of Biospheres* 32: 143 – 163.

Botta O., Martins Z. and Ehrenfreund P. 2007. Amino acids in Antarctic CM1 meteorites and their relationship to other carbonaceous chondrites. *Meteoritics and Planetary Science* 42: 81-92.

Brinton K. L. F., Engrand C., Glavin D. P., Bada J. L. and Maurette M. 1998. A search for extraterrestrial amino acids in carbonaceous Antarctic micrometeorites. *Origins of Life and Evolution of the Biosphere* 28: 413-424.

Busemann H., Young A. F., Alexander C. M. O'D., Hoppe P., Mukhopadhyay S. and Nittler L. R. 2006. Interstellar chemistry recorded in organic matter from primitive meteorites. *Science* 312: 727-730.

Chizmadia L. and Brearley A. J. 2003. Mineralogy and textural characteristics of fine-grained rims in the Yamato-791198 CM2 carbonaceous chondrite: Constraints on the location of aqueous alteration (abstract #1419). 34$^{th}$ Lunar and Planetary Science Conference. CD-ROM.

Clayton R. N. and Mayeda T. K. 1999. Oxygen isotope studies of carbonaceous chondrites. *Geochimica et Cosmochimica Acta* 63: 2089–2104.

Cody G. D. and Alexander C. M. O'D. 2005. NMR studies of chemical structural variations of insoluble organic matter from different carbonaceous chondrite groups. *Geochimica et Cosmochimica Acta* 69: 1085–1097.

Cronin J. R. and Chang S. 1993. Organic matter in meteorites: molecular and isotopic analyses of the Murchison meteorite. In *The chemistry of life's origin*, edited by Greenberg J. M., Mendoza-Gomez C. X., and Pirronello V. Dordrecht, The Netherlands: Kluwer Academic Publishing. pp. 209–258.

Cronin J. R., Pizzarello S. and Moore C. B. 1979. Amino acids in an Antarctic carbonaceous chondrite. *Science* 206: 335–337.

Cronin J. R., Cooper G. W. and Pizzarello S. 1995. Characteristics and formation of amino acids and hydroxy acids of the Murchison meteorite. *Advances in Space Research* 15: 91-97.

Ehrenfreund P., Glavin D. P., Botta O., Cooper G., and Bada J. L. 2001. Extraterrestrial amino acids in Orgueil and Ivuna: Tracing the parent body of CI type carbonaceous chondrites. *Proceedings of the National Academy of Sciences* 98: 2138–2141.

Glavin D. P., Matrajt G. and Bada J. L. 2004. Re-examination of amino acids in Antarctic micrometeorites. *Advances in Space Research* 33: 106-113.

Glavin D. P., Aubrey A., Dworkin J. P., Botta O. and Bada J. L. 2005. Amino acids in the Antarctic Martian meteorite MIL03346 (abstract #1920). 36th Lunar and Planetary Science Conference. CD-ROM.

Glavin D. P., Dworkin J. P., Aubrey A., Botta O., Doty III J. H., Martins Z. and Bada J. L. 2006. Amino acid analyses of Antarctic CM2 meteorites using liquid chromatography–time of flight–mass spectrometry. *Meteoritics and Planetary Science* 41: 889–902.

Grossman J. N. 1998. The Meteoritical Bulletin. *Meteoritics and Planetary Science* 33: A221-A239.

Grossman J. N. and Score R. 1996. The Meteoritical Bulletin. *Meteoritics and Planetary Science* 31: A161-A174.

Holzer G. and Oro J. 1979. The organic composition of the Allan Hills carbonaceous chondrite 77306 as determined by pyrolysis-gas chromatography-mass spectrometry and other methods. *Journal of Molecular Evolution* 13: 265-270.

Howe J. M., Featherstone W. R., Stadelman W. J and Banwartz G. J. 1965. Amino acid composition of certain bacterial cell-wall proteins. *Applied Microbiology* 13: 650-652.

Keller J.W., Baurick K. B., Rutt G. C., O'Malley M. V., Sonafrank N. L., Reynolds R. A., Ebbesson L. O. and Vajdos F. F. 1990. *Pseudomonas cepacia* 2,2-dialkylglycine decarboxylase. Sequence and expression in *Escherichia coli* of structural and repressor genes. *Journal of Biological Chemistry* 265: 5531-5539.

Kminek G., Botta O., Glavin D. P. and Bada J. L. 2002. Amino acids in the Tagish Lake meteorite. *Meteoritics and Planetary Science* 37: 697-701.

Kotra R. K., Shimoyama A., Ponnamperuma C. and Hare P. E. 1979. Amino acids in a carbonaceous chondrite from Antarctica. *Journal of Molecular Evolution* 13: 179-184.

Lerner N. R., Peterson E. and Chang S. 1993. The Strecker synthesis as a source of amino acids in carbonaceous chondrites - Deuterium retention during synthesis. *Geochimica et Cosmochimica Acta* 57: 4713–4723.

Martins Z., Watson J. S., Sephton M. A., Botta O., Ehrenfreund P. and Gilmour I. 2006. Free dicarboxylic and aromatic acids in the carbonaceous chondrites Murchison and Orgueil. *Meteoritics and Planetary Science* 41: 1073-1080.

Matrajt G., Pizzarello S., Taylor S. and Brownlee D. 2004. Concentration and variability of the AIB amino acid in polar micrometeorites: Implications for the exogenous delivery of amino acids to the primitive Earth. *Meteoritics and Planetary Science* 39: 1849-1858.

McDonald G. D. and Bada J. L. 1995. A search for endogenous amino acids in the Martian meteorite EETA 79001. *Geochimica et Cosmochimica Acta* 59: 1179-1184.

McSween H. Y. 1979. Alteration in CM carbonaceous chondrites inferred from modal and chemical variations in matrix. *Geochimica et Cosmochimica Acta* 43: 1761-1765.

O'Brien D. M., Fogel M. L. and Boggs C. L. 2002. Renewable and nonrenewable resources: Amino acid turnover and allocation to reproduction in Lepidoptera. *Proceedings of the National Academy of Sciences USA* 99: 4413-4418.

Peltzer E. T., Bada, J. L., Schlesinger G. and Miller S. L. 1984. The chemical conditions on the parent body of the Murchison meteorite: Some conclusions based on amino, hydroxy, and dicarboxylic acids. *Advances in Space Research* 4: 69–74.

Pizzarello S. and Garvie L. A. J. 2007. The organic composition of a CR2 chondrite: Differences and similarities with the Mighei-type meteorites (abstract #1393). 38[th] Lunar and Planetary Science Conference. CD-ROM.

Pizzarello S., Huang Y., Becker L., Poreda R. J., Nieman R. A., Cooper G. and Williams M. 2001. The organic content of the Tagish Lake meteorite. *Science* 293: 2236-2239.

Pizzarello S., Huang Y. and Fuller M. 2004. The carbon isotopic distribution of Murchison amino acids. *Geochimica et Cosmochimica Acta* 68: 4963-4969.

Scott J. H., O'Brien D. M., Fogel M. L., Emerson D., Sun H. and McDonald G. D. 2006. An examination of the carbon isotope effects associated with amino acid biosynthesis. *Astrobiology* 6: 867-880.

Sephton M. A. 2002. Organic compounds in carbonaceous meteorites. *Natural Product Reports* 19: 292–311.

Sephton M. A., Bland P. A., Pillinger C. T. and Gilmour I. 2004. The preservation state of organic matter in meteorites from Antarctica. *Meteoritics and Planetary Science* 39: 747-754.

Shimoyama A. and Harada K. 1984. Amino acid depleted carbonaceous chondrites (C2) from Antarctica. *Geochemical Journal* 18: 281–286.

Shimoyama A and Ogasawara R. 2002. Dipeptides and diketopiperazines in the Yamato-791198 and Murchison carbonaceous chondrites. *Origins of Life and Evolution of the Biosphere* 32: 165–179.

Shimoyama A., Ponnamperuma C. and Yanai, K. 1979. Amino Acids in the Yamato carbonaceous chondrite from Antarctica. *Nature* 282: 394-396.

Shimoyama A., Harada K and Yanai K. 1985. Amino acids from the Yamato-791198 carbonaceous chondrite from Antarctica. *Chemistry Letters* 8: 1183–1186.

Weisberg M. K. and Prinz M. 2000. The Grosvenor Mountains 95577 CR1 chondrite and hydration of the CR chondrites. *Meteoritics and Planetary Science* 35: A168.

Weisberg M. K., Prinz M., Clayton R. N. and Mayeda, T. K. 1993. The CR (Renazzo-type) carbonaceous chondrite group and its implications. *Geochimica et Cosmochimica Acta* 57: 1567-1586.

Zhao M., and Bada J. L. 1995. Determination of α-dialkylamino acids and their enantiomers in geological samples by high-performance liquid chromatography after derivatization with a chiral adduct of *o*-phthaldialdehyde. *Journal of Chromatography A* 690: 55-63.

**Table 1**. Summary of the average total amino acid abundances (in ppb) in the 6M HCl acid hydrolysed hot-water extracts of the EET92042, GRA95229 and GRO95577 meteorites measured by HPLC-FD*.

| Amino Acid | CR2 EET92042 | CR2 GRA95229 | CR1 GRO95577 |
|---|---|---|---|
| D-Aspartic acid | 467 ± 71 | 669 ± 7 | 13 ± 2 |
| L-Aspartic acid | 524 ± 76 | 696 ± 9 | 19 ± 4 |
| L-Glutamic acid | 3989 ± 97 | 3668 ± 319 | 40 ± 3 |
| D-Glutamic acid | 2309 ± 339 | 3005 ± 86 | 16 ± 6 |
| D,L-Serine† | 742 ± 42 | 1807 ± 84 | 50 ± 11 |
| Glycine | 26875 ± 1176 | 57796 ± 358 | 136 ± 14 |
| β-Alanine | 3005 ± 95 | 2910 ± 277 | 122 ± 6 |
| γ-ABA | 1975 ± 176 | 2848 ± 146 | 54 ± 6 |
| DL-β-AIB†‡ | 1526 ± 88 | 1645 ± 61 | 30 ± 2 |
| D-Alanine | 23862 ± 324 | 50722 ± 419 | 74 ± 22 |
| L-Alanine | 23215 ± 609 | 50681 ± 2884 | 96 ± 20 |
| DL-β-ABA† | 3094 ± 149 | 5986 ± 83 | 49 ± 5 |
| α-AIB | 57856 ± 2030 | 27679 ± 1113 | 48 ± 3 |
| D, L-Isovaline | ≤22798§ | ≤27844§ | ≤131§ |
| L-Valine | 3632 ± 60 | 6053 ± 150 | 13 ± 4 |
| D-Valine | 3665 ± 92 | 5736 ± 205 | 8 ± 3 |
| **Total** | **180000** | **249000** | **900** |

*Quantification of the amino acids included background level correction using a serpentine blank. The associated errors are based on the standard deviation of the average value between six separate measurements (N) with a standard error, $\delta x = \sigma_x \cdot N^{-1/2}$
†Enantiomers could not be separated under the chromatographic conditions.
‡Optically pure standard not available for enantiomeric identification.
§These values are upper limits because there is the possibility of co-elution with α-ABA.

**Table 2**. Summary of the average total amino acid abundances (in ppb) in the 6M HCl acid hydrolysed hot-water extracts of the EET92042, GRA95229 and GRO95577 meteorites measured by GC-MS*.

| Amino Acid | EET92042 | GRA95229 |
|---|---|---|
| D-Aspartic acid | 409 ± 41 | 551 ± 75 |
| L-Aspartic acid | 465 ± 24 | 576 ± 51 |
| L-Glutamic acid | 4468 ± 503 | 4209 ± 415 |
| D-Glutamic acid | 3090 ± 422 | 3489 ± 389 |
| Glycine | 24975 ± 608 | 40496 ± 1028 |
| β-Alanine | 3046 ± 50 | 3143 ± 495 |
| γ-ABA | 1512 ± 66 | 1914 ± 398 |
| DL-β-AIB‡ | 1429 ± 333 | 2091 ± 405 |
| D-Alanine | 21664 ± 1009 | 52465 ± 6860 |
| L-Alanine | 22297 ± 1583 | 51141 ± 6272 |
| D-β-ABA | 1327 ± 33 | 3903 ± 377 |
| L-β-ABA | 1458 ± 99 | 4239 ± 494 |
| α-AIB | 50210 ± 870 | 30257 ± 1226 |
| D,L-Isovaline† | 22806 ± 459 | 29245 ± 2229 |
| L-Valine | 2084 ± 129 | 6996 ± 700 |
| D-Valine | 1969 ± 255 | 7154 ± 788 |
| D-α-ABA | 1123 ± 54 | 2956 ± 125 |
| L-α-ABA | 1244 ± 28 | 2955 ± 120 |
| **Total** | **165000** | **247300** |

*Quantification of the amino acids included background level correction using a serpentine blank.
†Enantiomers could not be separated under the chromatographic conditions.
‡Optically pure standard not available for enantiomeric identification.

**Table 3.** Amino acid enantiomeric ratios (D/L) in the CR carbonaceous chondrites EET92042, GRA95229 and GRO95577*.

| Amino Acids | CR2 EET92042[†] | CR2 EET92042[‡] | CR2 GRA95229[†] | CR2 GRA95229[‡] | CR1 GRO95577[†] |
|---|---|---|---|---|---|
| Aspartic acid | 0.89 ± 0.19 | 0.88 ± 0.10 | 0.96 ± 0.02 | 0.96 ± 0.16 | 0.68 ± 0.18 |
| Glutamic acid | 0.58 ± 0.09 | 0.69 ± 0.12 | 0.82 ± 0.08 | 0.83 ± 0.12 | 0.40 ± 0.15 |
| Alanine | 1.03 ± 0.03 | 0.97 ± 0.08 | 1.00 ± 0.06 | 1.03 ± 0.18 | 0.77 ± 0.28 |
| β-ABA | ¶ | 0.91 ± 0.07 | ¶ | 0.92 ± 0.14 | ¶ |
| Valine | 1.01 ± 0.03 | 0.94 ± 0.14 | 0.95 ± 0.04 | 1.02 ± 0.15 | 0.62 ± 0.30 |
| α-ABA | ¶ | 0.90 ± 0.05 | ¶ | 1.00 ± 0.06 | ¶ |

*The uncertainties are based on the absolute errors shown in Tables 1 and 2, and are obtained by standard propagation calculations.
[†]D/L ratios calculated from the concentrations reported in Table 1, measured by HPLC-FD.
[‡]D/L ratios calculated from the concentrations reported in Table 2, measured by GC-MS.
[¶]Not determined, because enantiomeric separation was not possible or amino acid abundance was not determined.

**Table 4**. Summary of the $\delta^{13}C$ values (‰) of amino acids in the EET92041 and GRA95229 meteorites*.

| Amino Acid | EET92042 | GRA95229 |
|---|---|---|
| D-Asp. acid | +34.4 ± 4.1 | +34.9 ± 0.5 |
| L-Asp. acid | +23.4 ± 0.7 | +33.0 ± 3.1 |
| L-Glu. acid | -19.5 ± 1.7 | -17.6 ± 1.9 |
| D-Glu. acid | +46.1 ± 2.1 | +47.2[‡] |
| Glycine | +31.8 ± 2.0 | +33.8 ± 1.6 |
| D-Alanine | +44.5 ± 2.0 | +41.7 ± 2.4 |
| L-Alanine | +49.9[‡] | +40.9 ± 6.2 |
| α-AIB | § | +31.6 ± 6.1 |
| Isovaline[†] | § | +50.5[‡] |

*The associated errors are based on the standard deviation of the average value between three and five separate measurements (N) with a standard error, $\delta x = \sigma_x \cdot N^{-1/2}$
[†]Enantiomers could not be separated under the chromatographic conditions.
[‡]Average of two repeated analyses.
[§]Not determined.

# Figure Legends

**Fig. 1**. The 0 to 40 min region (no peaks were observed outside this region) of the HPLC-FD chromatograms. OPA/NAC derivatization (1 minute) of amino acids in: (A) the standard, the 6M HCl-hydrolyzed hot-water extracts from the CR2 carbonaceous chondrite EET92042 and GRA95229, and the serpentine blank; (B) the 6M HCl-hydrolyzed hot-water extracts from the CR1 carbonaceous chondrite GRO95577 and corresponding serpentine blank. HPLC-FD chromatograms (A) and (B) are not on the same scale and were not run on the same day. Peaks were identified by comparison of the retention time to those in the amino acid standard run on the same day: 1. D-Aspartic acid; 2. L-Aspartic acid; 3. L-Glutamic acid; 4. D-Glutamic acid; 5. D, L-Serine; X. Unknown; 6. Glycine; 7. β-Alanine; 8. γ-ABA; 9. D, L-β-AIB; 10. D-Alanine; 11. L-Alanine; 12. D, L-β-ABA; 13. α-AIB; 14. D-Isovaline; 15. L-Isovaline; 16. L-Valine; 17. D-Valine.

**Fig. 2**. Single ion GC-MS traces (*m/z* 69, 126, 138, 140, 154, 168, 180, 182, and 184) of the derivatized (*N*-TFA, *O*-isopropyl) EET92042, GRA95229 and serpentine blank HCl-hydrolyzed hot-water extracts, and amino acid standard. The peaks were identified by comparison of the retention time and mass fragmentation pattern to those in the amino acid standard run on the same day: 1. α-AIB; 2. Isovaline; 3. D-Alanine; 4. L-Alanine; 5. D-α-ABA; 6. L-α-ABA+D-Valine; 7. L-Valine; 8.Glycine; 9. β-AIB; 10. β-Alanine; 11. D-β-ABA; 12. L-β-ABA; 13. γ-ABA; 14. D-Aspartic acid; 15. L-Aspartic acid; 16. D-Glutamic acid; 17. L-Glutamic acid.

**Fig. 3**. (A) Total amino acid abundances (in ppb) for the α-amino acids (glycine, alanine, α-ABA, α-AIB, isovaline and valine), β-amino acids (β-alanine, β-ABA and β-AIB), γ-amino acid (γ-ABA) and dioic amino acids (aspartic acid and glutamic acid) present in the CR2s GRA95229 (▲) and EET92042 (■) (this work; data taken from Table 1 and 2), the CM2 Y791198 (Δ) (Shimoyama et al. 1985; Shimoyama and Ogasawara 2002), the CR2 Renazzo (━) (Botta et al. 2002), the CR1 GRO95577 (O) (this work; data taken from Table 1) and the CM1 LAP02277 (*) (Botta et al. 2007). Straight and branched carbon chain amino acids plotted by increasing carbon number, respectively. In the case of the Renazzo and LAP02277 meteorites, not all amino acid abundances are available from the literature (Botta et al. 2002; Botta et al. 2007). The abundance of α-ABA was not determined for the GRO95577 meteorite (this work, Table 1). (B) Relative amino acid abundances (glycine = 1) for the amino acids α-aminoisobutyric acid (stripes), β-alanine (white), γ-ABA (gray), aspartic acid (black) and glutamic acid (dots) in the CR2s GRA95229 and EET92042 (this work; data taken from Table 1 and 2), the CM2 Y791198 (Shimoyama et al. 1985; Shimoyama and Ogasawara 2002), the CR2 Renazzo (Botta et al. 2002), the CR1 GRO95577 (this work; data taken from Table 1) and the CM1 LAP02277 (Botta et al. 2007).

**Fig. 4.** Typical GC-C-IRMS chromatogram obtained in this study. (A) *m/z* 44 trace (bottom) and ratio between the *m/z* 45 and *m/z* 44 trace (top) for the GC-C-IRMS analysis of a portion of the GRA95229 HCl-hydrolyzed hot-water extract containing the α-amino acids 1. α-AIB; 2. Isovaline; 3. D-Alanine; 4. L-Alanine; 5. Glycine. (B) *m/z* 44 trace

(bottom) and ratio between the *m/z* 45 and *m/z* 44 trace (top) for the GC-C-IRMS analysis of a portion of the GRA95229 HCl-hydrolyzed hot-water extract containing the following amino acids: 6. D-Aspartic acid; 7. L-Aspartic acid; 8. D-Glutamic acid; 9. L-Glutamic acid.

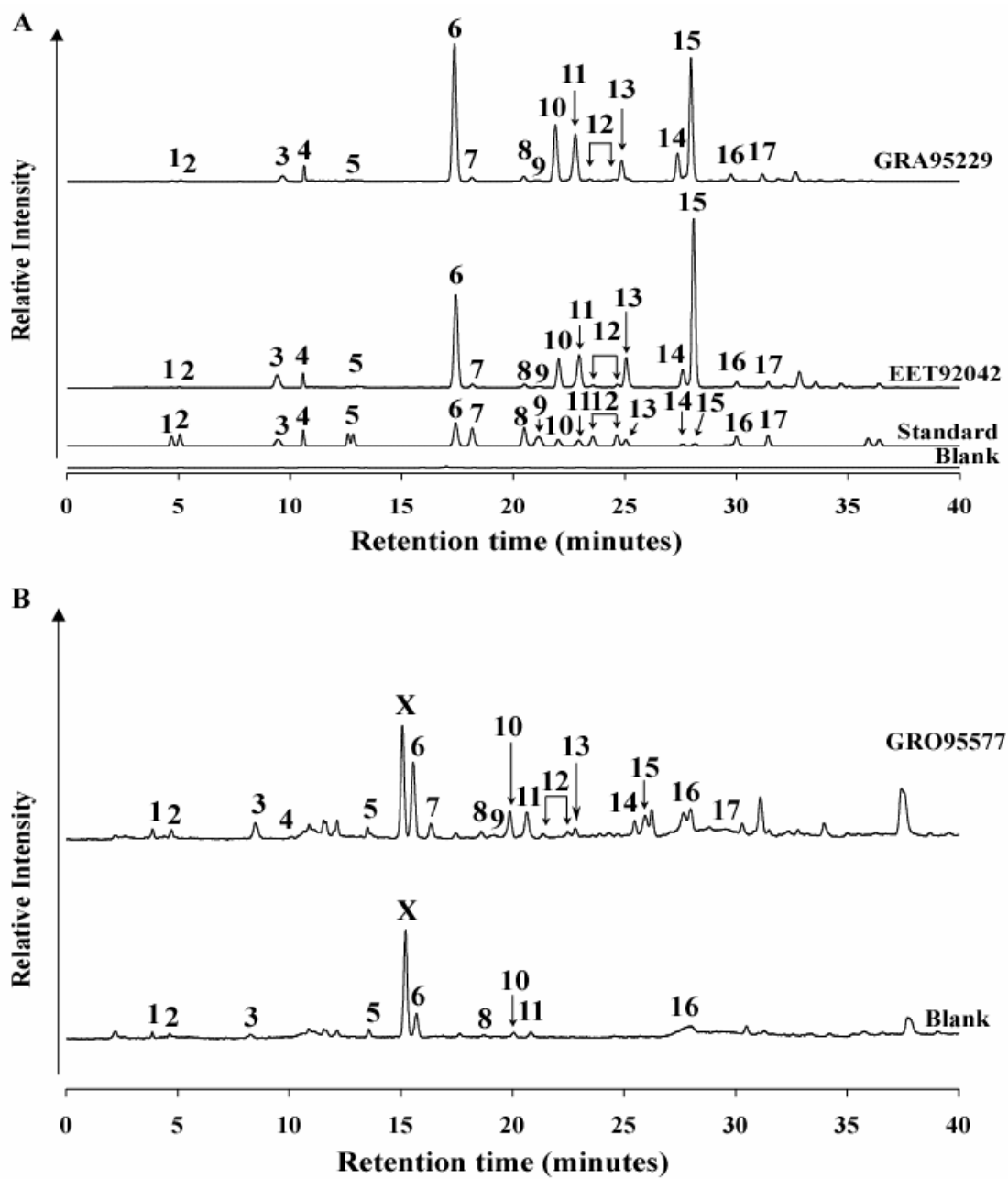

Fig. 1

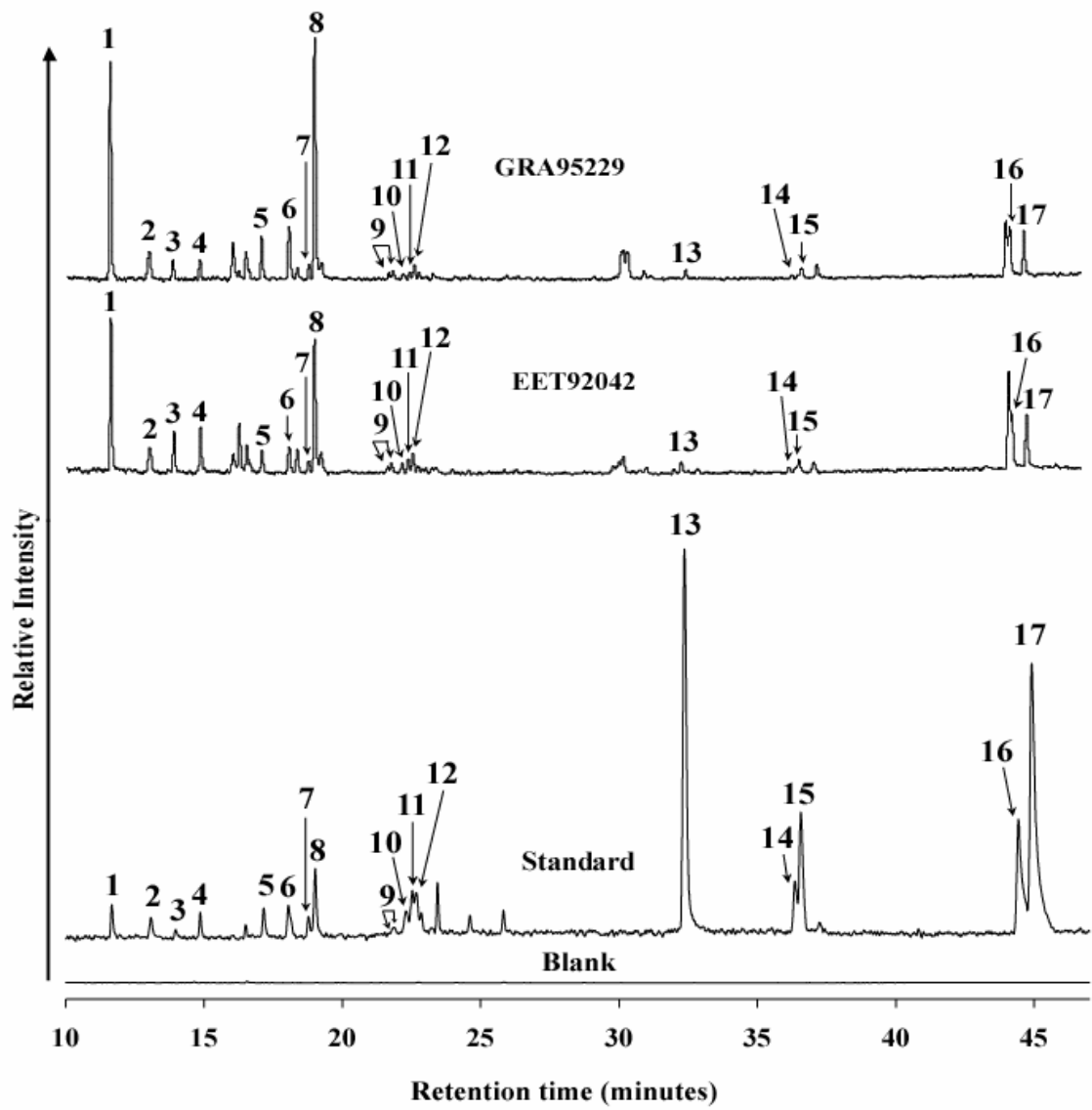

**Fig. 2**

**Fig. 3**

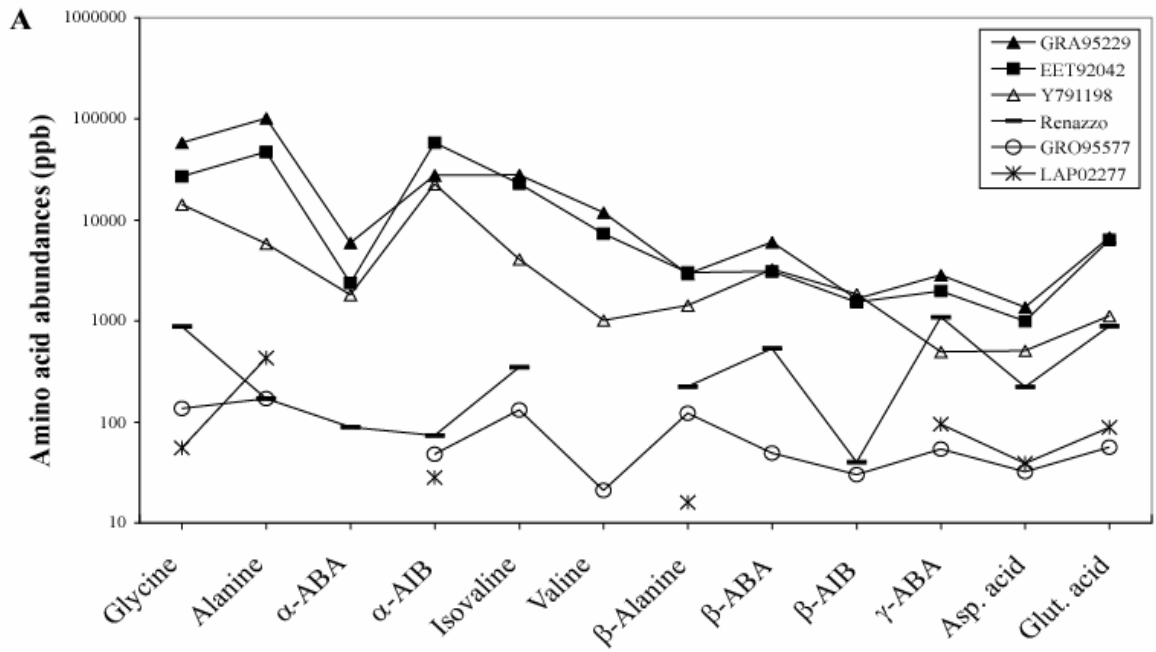

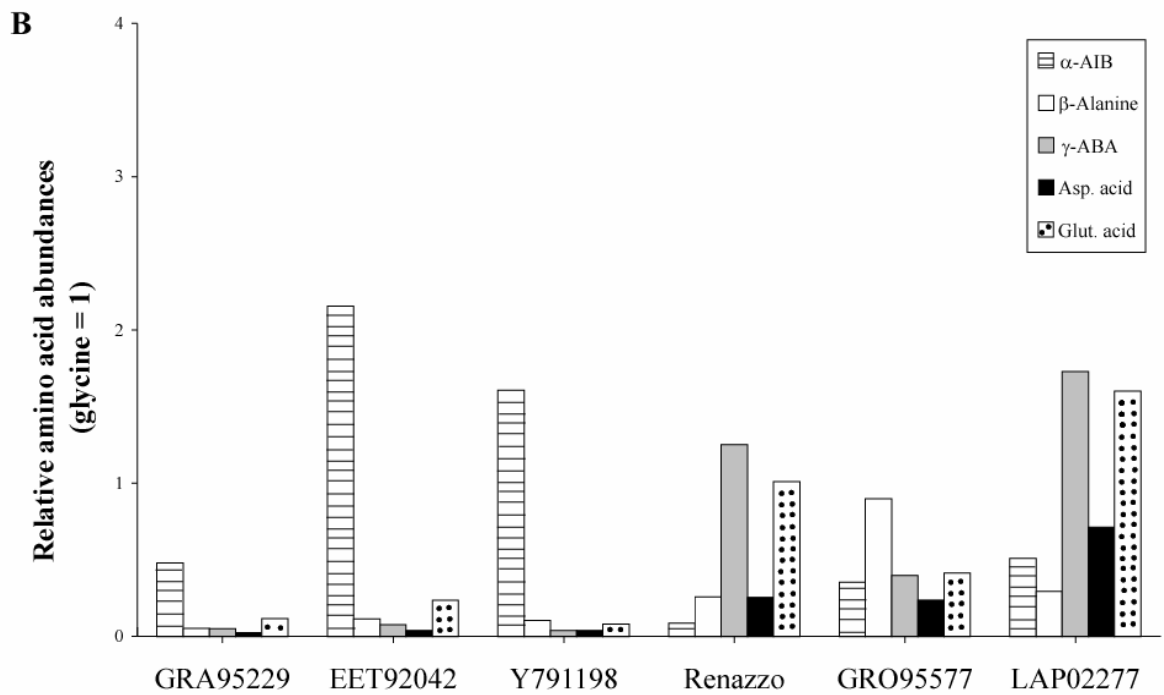

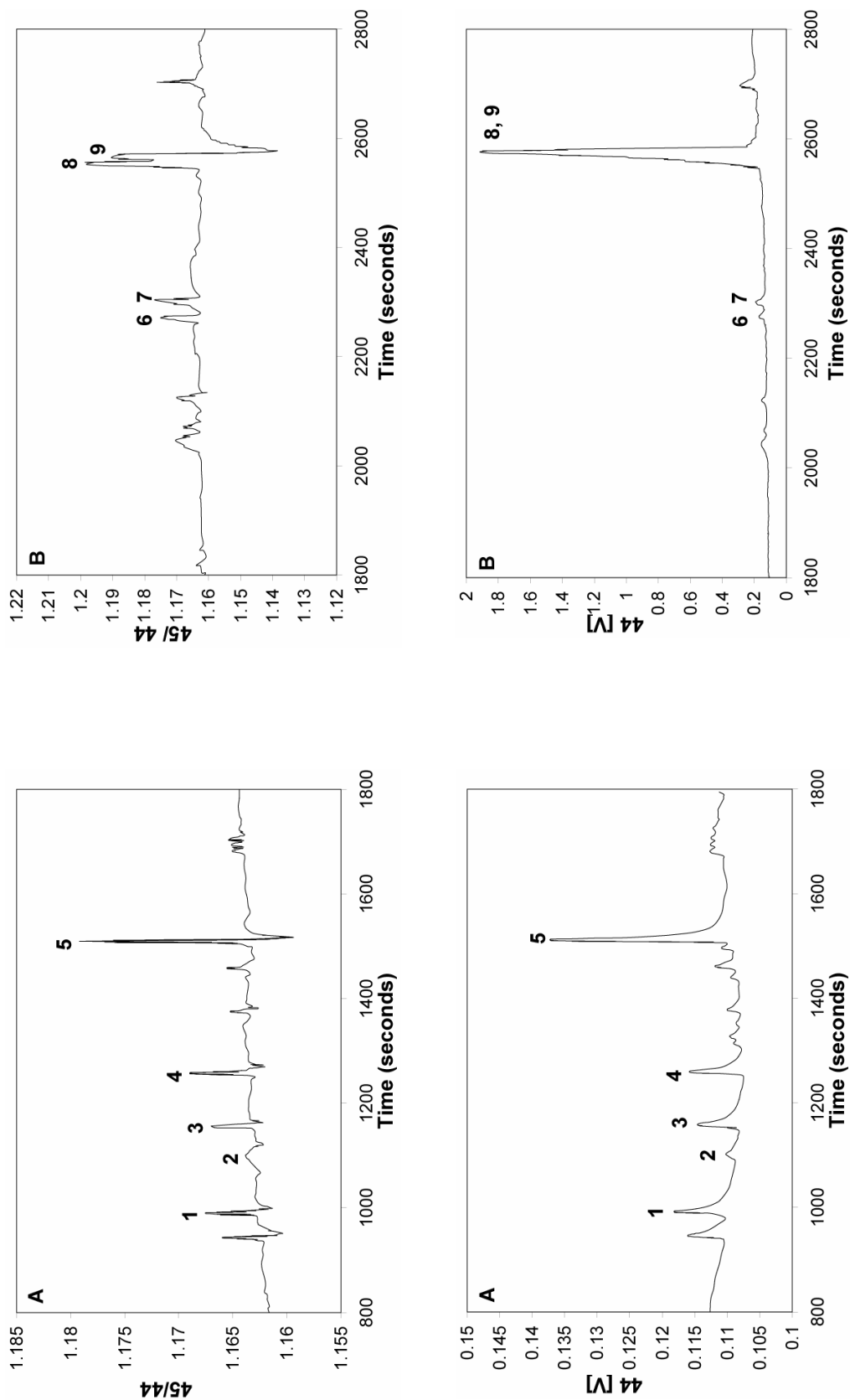

Fig. 4